\documentclass[a4paper,fleqn,usenatbib]{mnras}

\usepackage[T1]{fontenc}
\usepackage{ae,aecompl}

\usepackage{graphicx}	% Including figure files
\usepackage{amsmath}	% Advanced maths commands
\usepackage{amssymb}	% Extra maths symbols
\usepackage{subfigure}

\newcommand{\dinamo}{{\sc dinamo} }
\newcommand{\pcc}{\,{\rm cm}^{-3}}

\newcommand{\um}{\, {\rm \mu m}}

\newcommand{\kel}{\, {\rm K}}

\newcommand{\msun}{\, {\rm M}_\odot}

\newcommand{\nel}{n_{\rm e}}

\newcommand{\flux}{\, {\rm erg} \, {\rm cm}^{-2} \, {\rm s}^{-1}}
\newcommand{\pc}{\, {\rm pc}}
\newcommand{\amin}{a_{\rm min}}
\newcommand{\amax}{a_{\rm max}}
\newcommand{\ssum}{\displaystyle\sum}
\newcommand{\hv}{h \nu}
\newcommand{\nion}{n_{\rm i}}
\newcommand{\tel}{T_{\rm e}}
\newcommand{\tion}{T_{\rm i}}
\newcommand{\mh}{m_{\rm H}}
\newcommand{\ev}{\, {\rm eV}}
\newcommand{\kpc}{\, {\rm kpc}}

\title[Dust in the Cas A SNR]{The mass, location and heating of the dust in the Cassiopeia A supernova remnant}

\author[F. D. Priestley et al.]{
F. D. Priestley$^{1}$
M. J. Barlow$^{1}$
and I. De Looze$^{1,2}$
\\
$^{1}$Department of Physics and Astronomy, University College London, Gower Street, London WC1E 6BT, UK\\
$^{2}$Sterrenkundig Observatorium, Ghent University, Krijgslaan 281 - S9, 9000 Gent, Belgium\\
}

\date{Accepted XXX. Received YYY; in original form ZZZ}

\pubyear{2018}

\begin{document}
\label{firstpage}
\pagerange{\pageref{firstpage}--\pageref{lastpage}}
\maketitle

\begin{abstract}
We model the thermal dust emission from dust grains heated by synchrotron radiation and by particle collisions, under conditions appropriate for four different shocked and unshocked gas components of the Cassiopeia A (Cas A) supernova remnant (SNR). By fitting the resulting spectral energy distributions (SEDs) to the observed SNR dust fluxes, we determine the required mass of dust in each component. We find the observed SED can be reproduced by $\sim 0.6 \msun$ of silicate grains, the majority of which is in the unshocked ejecta and heated by the synchrotron radiation field. Warmer dust, located in the X-ray emitting reverse shock and blastwave regions, contribute to the shorter wavelength infrared emission but make only a small fraction of the total dust mass. Carbon grains can at most make up $\sim 25 \%$ of the total dust mass. Combined with estimates for the gas masses, we obtain { dust-to-gas} mass ratios for each component, which suggest that the condensation efficiency in the ejecta is high, and that dust in the shocked ejecta clumps is well protected from destruction by sputtering in the reverse shock.
\end{abstract}


%%%%%%%%%%%%%%%%%%%%%%%%%%%%%%%%%%%%%%%%%%%%%%%%%%

%%%%%%%%%%%%%%%%% BODY OF PAPER %%%%%%%%%%%%%%%%%%

\section{Introduction}

The detections of significant ($\gtrsim 10^8 \msun$) masses of dust in high-redshift quasars \citep{bertoldi2003,priddey2003}, and dust-enriched galaxies at redshifts $z > 7$ \citep{watson2015,laporte2017,hashimoto2018}, require an explanation of how sufficient quantities of dust can be formed at such early epochs. Core-collapse supernovae (SNe) have been proposed as a potential source of this dust \citep{dunne2003,gall2011}, as their progenitors evolve rapidly compared to the age of the universe at these redshifts ($\sim 1 \, {\rm Gyr}$). Observations of supernova remnants (SNRs) have confirmed the presence of large quantities of dust formed in the ejecta, both through infrared (IR)/sub-millimetre (submm) detection of dust emission \citep{barlow2010,matsuura2011,gomez2012,matsuura2015,delooze2017} and alteration of emission line profiles due to dust extinction \citep{bevan2016,bevan2017}.

In order to explain the observed dust masses at high redshift, the average dust yield per SNe must exceed some minimum value, estimated as $\sim 1 \msun$ by \citet{dwek2007} and between $0.1-1 \msun$ by \citet{morgan2003} and \citet{michalowski2010}, although \citet{rowlands2014} found that higher yields may be necessary if dust destruction in the interstellar medium (ISM) is taken into account. Dust masses observed in SNRs, such as the Crab Nebula ($0.1-0.2 \msun$; \citet{gomez2012}) and SN 1987A ($0.8 \msun$; \citet{matsuura2015}), approach or exceed this value, but the fraction which will survive passage through the SN reverse shock and into the ISM is uncertain \citep{nozawa2007,bianchi2007,nozawa2010,micelotta2016,biscaro2016,bocchio2016}. In particular, large ($a \gtrsim 0.1 \um$) dust grains are able to survive destruction by sputtering much more effectively than smaller grains \citep{silvia2010}.

Cassiopeia A (Cas A) is a Galactic SNR located $3.4 \, {\rm kpc}$ away \citep{reed1995}, with an age of approximately $330 \, {\rm yr}$ \citep{fesen2006} and a radius of $1.7 \pc$ \citep{reed1995}. It provides a unique laboratory to test the efficiency of dust condensation in SN ejecta, and the subsequent destruction of dust by the reverse shock, and as such has been studied extensively in the past. IR/submm observations have led to derived dust masses ranging from $\sim 10^{-4} \msun$ of hot ($T \sim 100 \kel$) dust \citep{arendt1999,douvion2001} to $2-4 \msun$ of cold dust emitting at sub-mm wavelengths \citep{dunne2003}, although this higher mass has been attributed to foreground dust emission in the ISM \citep{krause2004}. Analyses of integrated fluxes from {\it Spitzer} and {\it Herschel} observations \citep{rho2008,barlow2010,arendt2014} found $\sim 0.01 \msun$ of hot dust, with $\sim 0.1 \msun$ of cold, unshocked dust present in the central regions, in agreement with simulations of the dust formation and evolution in Cas A by \citet{nozawa2010}. \citet{dunne2009} suggested the observed polarization of the submm emission could be explained by $\sim 1 \msun$ of cold dust, similar to the value of $1.1 \msun$ given by \citet{bevan2017} as the most likely mass based on the shape of emission line profiles affected by extinction. \citet{delooze2017} utilised spatially resolved {\it Herschel} and {\it Spitzer} observations of Cas A to fit the dust continuum emission, following the removal of line and synchrotron contamination, using a four-component model including ISM dust emission and three SNR dust temperature components. They found a large mass of unshocked cold dust in the centre of the SNR (up to $0.6 \msun$), significantly above previous estimates based on the IR/submm emission.

Previous modelling of the Cas A dust emission has been based on fitting the spectral energy distribution (SED) with some number of temperature components for a given dust composition (i.e. `hot' and `cold' dust). This assumes all dust grains radiate at the same temperature for each component, but grains of different sizes will in general have different equilibrium temperatures for the same heating source. Additionally, smaller grains can undergo large temperature fluctuations (e.g. \citealt{purcell1976,draine1985,dwek1986}) and may not reach an equilibrium temperature at all. \citet{temim2013} found that modelling the emission from a distribution of grain sizes in the Crab Nebula reduced the dust mass estimate by a factor of two compared to two-temperature fits by \citet{gomez2012}, demonstating the importance of accounting for these effects. In this paper, we calculate the emission from a population of grains subjected to conditions appropriate for the various gas components in Cas A, and use it to constrain the mass, properties and location of the newly-formed dust.

\section{Physical properties of the Cas A SNR}
\label{sec:prop}

In order to determine the dust emission from the remnant, several physical properties are required: the densities and temperatures of the electrons and nuclei, the dominant type of nucleus, and the radiation field strength and spectrum. Observations of Cas A reveal a complex structure, with material covering a wide range of densities and temperatures emitting at different wavelengths. The supernova explosion has driven a forward shock into the circumstellar material, thought to be from the stellar wind of the progenitor \citep{hwang2009}, while the ejecta from the supernova itself crosses the reverse shock as it expands \citep{delaney2004}. Both shocks are visible as X-ray emitting regions, with typical densities of $n \sim 1-10 \pcc$ and temperatures $T \gtrsim 10^7 \kel$ \citep{willingale2003,lazendic2006,patnaude2014,wang2016}. The ejecta is mostly comprised of heavy elements, principally oxygen \citep{chevalier1979,willingale2003}. As well as the X-ray emitting gas, the shocked ejecta also consists of denser clumps or knots, emitting in the optical and IR \citep{hurford1996,delaney2010,patnaude2014} and associated with the dust emission \citep{arendt1999}. Electron densities in the shocked clumps are $\nel \sim 10^3-10^5\pcc$ \citep{smith2009,delaney2010,lee2017}, while the gas temperatures are of order $10^4 \kel$ \citep{arendt1999,docenko2010}. The SNR also contains ejecta which has not yet encountered the reverse shock, and is consequently much cooler. \citet{smith2009} estimated a maximum electron density of $\nel \lesssim 100 \pcc$ for the unshocked ejecta based on forbidden line ratios, while observations of radio absorption by \citet{delaney2014} and \citet{arias2018} give $\nel \sim 10 \pcc$ and $T \sim 100 \kel$. \citet{raymond2018} inferred a preshock temperature of $\sim 100 \kel$ from [Si I] IR emission lines.

\citet{krause2008} determined that the Cas A SN was of type IIb from a spectrum of its light echo, meaning that the progenitor star must have lost most of its hydrogen envelope pre-explosion. \citet{young2006} suggested a progenitor with main-sequence mass of $15-25 \msun$ and a mass at explosion of $4-6 \msun$, based on a comparison of stellar evolution and explosion models with observed features of the SNR. Modelling of the X-ray spectra \citep{vink1996,willingale2003} gives ejecta masses in the $2-4 \msun$ range, with the swept-up material in the forward shock contributing an additional $8-10 \msun$. Based on observed emission line strengths from \citet{smith2009}, \citet{docenko2010} and \citet{milisavljevic2013} (see Appendix \ref{sec:clump}), the optical and IR-emitting knots have a total gas mass of $0.59 \msun$. \citet{arias2018} estimate an unshocked ejecta gas mass of $\sim 3 \msun$. They noted that this conflicts with some models of the emission from the shocked regions, which suggest most of the ejecta has already passed the reverse shock \citep{chevalier2003,laming2003}, and that their result is sensitive to both the assumed gas temperature and whether the gas is clumped. However, most of the dust ($0.4-0.6 \msun$), whose mass was derived by \citet{delooze2017}, appears to be located inside the reverse shock. The gas temperature has since been measured to be $\sim 100 \kel$ \citep{raymond2018}, the value used by \citet{arias2018} to give the $3 \msun$ mass estimate, although the (unknown) degree of clumping still allows for a potentially lower unshocked ejecta mass.

{ The forward and reverse shocks generate synchrotron radiation, at wavelengths ranging from radio to X-ray (e.g. \citealt{hwang2004,wang2016}). The low-frequency part of the synchrotron SED can be fit by a single power law \citep{delooze2017}, but this overestimates the flux at the higher frequencies important for dust heating. We find that a power law with an exponential cutoff, of the form $\nu F_\nu \propto \nu^{0.2} \exp(-(\nu/\nu_0)^{-0.4})$ with $\nu_0 = 900 \ev$, provides a good fit to the radio and X-ray data from \citet{wang2016}, similar to theoretical predictions from \citet{zirakashvili2007}. The total synchrotron luminosity, assuming a distance of $3.4 \kpc$ \citep{reed1995}, is $8 \times 10^{36} \, {\rm erg} \, {\rm s}^{-1}$.}

\section{Method}

\subsection{DINAMO}

\begin{figure}
  \centering
  \includegraphics[width=\columnwidth]{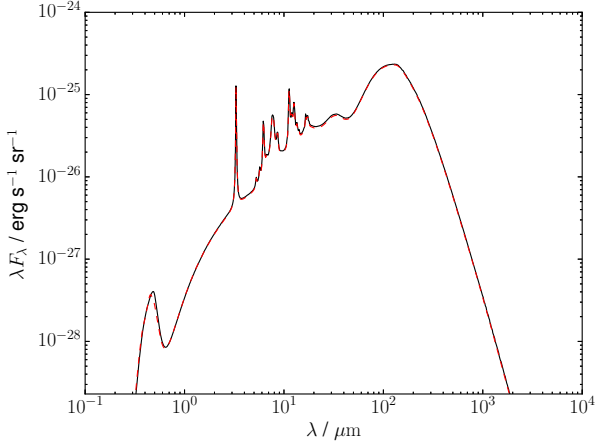}
  \caption{Dust emission for the \citet{mathis1983} ISM radiation field benchmark case from \citet{camps2015}, as calculated by \dinamo (solid black line) and DustEM (dashed red line).}
  \label{fig:casA:benchmark}
\end{figure}

We make use of a new dust emission code, \dinamo (\dinamo Is Not A MOdified blackbody)\footnote{\url{https://fpriestley.github.io/dinamo}}. The code calculates the equilibrium temperature distributions and the thermal emission for a population of dust grains of arbitrary sizes and compositions, given the physical properties of the environment the grains are located in (radiation field, electron/ion density and temperature). \dinamo includes heating from both radiation and particle collisions simultaneously, and treats cooling as a state-to-state process rather than using the continuous cooling approximation \citep{siebenmorgen1992}. The code has been benchmarked against DustEM \citep{compiegne2011} using the test cases from \citet{camps2015}, and found to be in excellent agreement, as shown in Figure \ref{fig:casA:benchmark}.

The stochastic heating of dust grains is treated following the method of \citet{guhathakurta1989}. For each grain size, $N$ enthalpy bins are defined between a maximum and minimum enthalpy, with the probability of a grain being found in bin $i$ at time $t$ given by $P_i(t)$. The probability per unit time of a grain moving from bin $i$ to bin $f$ is $A_{fi}$, with $A_{ii} = - \ssum_{f \ne i} A_{fi}$, so that $\frac{d}{dt}P_f = \ssum_i A_{fi} P_i$. For the steady state solution, $\ssum_i A_{fi} P_i = 0$ for all $i$, which with the additional constraint that $\ssum P_i = 1$ gives a system of linear equations which can be solved to find the `equilibrium' temperature distribution for a particular grain size and species. The details of the implementation of radiative and collisional heating, and cooling by thermal emission, are given in Appendix \ref{sec:rates}.

\subsection{Input parameters}

\begin{figure}
  \centering
  \includegraphics[width=\columnwidth]{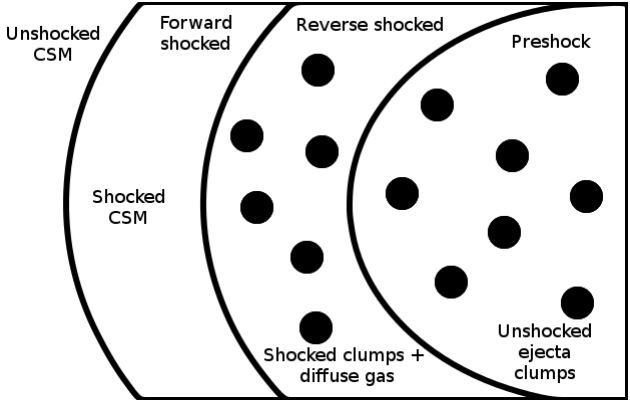}
  \caption{{ Schematic diagram representing the assumed location of each component with respect to the forward and reverse shocks. The unshocked ejecta consists of clumped material towards the right of the diagram. Moving left, the ejecta encounters the reverse shock, compressing and heating the clumps, and ablating material which becomes the diffuse component. Further left is a contact discontinuity between the shocked ejecta and shocked circumstellar material (CSM) swept up by the blast wave of the forward shock, and the unshocked CSM which the forward shock is propagating into.}}
  \label{fig:schematic}
\end{figure}

Based on the above discussion, we model the SNR as consisting of four main components - the unshocked, { clumped} ejecta, the (reverse) shocked ejecta clumps responsible for the optical emission, the X-ray emitting diffuse reverse shocked ejecta and the { X-ray emitting} material swept up by the blast wave. { We show a schematic diagram of the location of these components in Figure \ref{fig:schematic}.} The adopted densities and temperatures of these components are listed in Table \ref{tab:casA:gasprop}. For the unshocked ejecta we take $\nel = 100 \pcc$ and $T = 100 \kel$, the upper limits from observations \citep{smith2009,raymond2018} - we show later that lower values have no effect on the resulting emission, as the heating is dominated by the radiation field. { We assume an ionization fraction of unity for convenience, whereas at these temperatures the gas is presumably mostly neutral. \citet{arias2018} find $\nel \sim 10 \pcc$ for the unshocked ejecta - for a nucleon density of $\nion = 100 \pcc$, this gives an ionization fraction of $0.1$, not unreasonable for material irradiated by UV and X-ray photons and with an enhanced abundance of low ionization potential elements (e.g. Si, Fe).} For the shocked clumps we use $\nel = 480 \pcc$, used in deriving the total gas mass in this component, and $T = 10^4 \kel$ \citep{arendt1999,docenko2010}. { From the ionization states of oxygen (Appendix \ref{sec:clump}) in the shocked clumps, the ionization fraction must be approximately unity.} As the values of $\nel$ derived from observations range from $10^2-10^5 \pcc$, we also consider models with higher values of this parameter. For the two X-ray emitting components we use the electron and ion temperatures and densities from \citet{willingale2003}. For the other two components we assume that $\tion = \tel$ and $\nion = \nel$ - electrons are much more efficient at heating than ions for the same temperature due to their lower mass and correspondingly higher thermal velocities, so the exact value of $\nion$ in these cases is unlikely to be significant. In the three ejecta components we assume the ionic species is oxygen, using the heating efficiency from \citet{dwek1987}, while in the forward shock we use hydrogen. \citet{willingale2003} found mean ionic masses of $15.6$ and $1.33 \mh$ for the reverse and forward shocks respectively, so oxygen (ejecta) and hydrogen (forward shock) are almost certainly the dominant constituents.

\begin{table*}
  \centering
  \caption{Adopted gas masses, ion and electron number densities and temperatures, and dominant ionic species for the four gas components. References are (1) \citet{smith2009} (2) \citet{raymond2018} (3) \citet{arias2018} (4) \citet{docenko2010} (5) Appendix \ref{sec:clump} (6) \citet{willingale2003}.}
  \begin{tabular}{cccccccc}
    \hline
    Component & $M_{\rm gas} / \msun$ & $\nion / \pcc$ & $\nel / \pcc$ & $\tion / \kel$ & $\tel / \kel$ & Ion & Ref. \\
    \hline
    Preshock & $3$ & $100$ & $100$ & $100$ & $100$ & O & (1),(2),(3) \\
    Clumped & $0.59$ & $480$ & $480$ & $10^4$ & $10^4$ & O & (4),(5) \\
    Diffuse & $1.68$ & $7.8$ & $61$ & $7.05 \times 10^8$ & $5.22 \times 10^6$ & O & (6) \\
    Blastwave & $8.32$ & $14.3$ & $16$ & $3.98 \times 10^8$ & $3.79 \times 10^7$ & H & (6) \\ \hline
  \end{tabular}
  \label{tab:casA:gasprop}
\end{table*}

{ In addition to heating by the gas, dust in Cas A is also heated by the ambient radiation field. In Section \ref{sec:prop} we determined the luminosity and SED of the synchrotron radiation emitted by the forward and reverse shocks, but the local intensity which enters the calculation of the dust heating depends on both the location of the dust, and the distribution of the emitting material. Based on an SNR radius of $1.7 \pc$ \citep{reed1995}, we assume all components are located $1 \pc$ from a source with the luminosity of the whole remnant, and explore the sensitivity of our results to changes in the intensity.}

Given the oxygen-rich nature of the SNR, the dust in Cas A is expected to be primarily composed of silicates. \citet{rho2008} and \citet{arendt2014} found magnesium silicates of various compositions could reproduce a strong $21 \um$ feature in the {\it Spitzer} dust emission spectra, although Al$_2$O$_3$, carbon grains and other species were also suggested to be present. We use optical constants for magnesium and magnesium-iron silicates with varying elemental ratios \citep{jaeger1994,dorschner1995,jaeger2003}, which span the range $0.2-500 \um$. As these do not extend into the shorter-wavelength regions important for dust heating, we use the optical constants for astronomical silicates from \citet{laor1993} for $0.001-0.2 \um$ for all silicate species, interpolating between the two data sets to avoid discontinuities, and we extrapolate the experimental data up to $1000 \um$. We also investigated carbon grains, using optical constants for the ACAR and BE samples from \citet{zubko1996}, extended to $0.0003 \um$ with data from \citet{uspenskii2006}, as described by \citet{owen2015}. We assume mass densities of $2.5$ and $1.6 \, {\rm g} \pcc$ for silicate and carbon grains respectively, following \citet{delooze2017}, and sublimation temperatures of $1500$ and $2500 \kel$, although our results are not sensitive to the choice of this parameter. Dust properties used are summarised in Table ~\ref{tab:casA:dustprop}.

\begin{table}
  \centering
  \caption{Dust species and their adopted densities $\rho_g$, sublimation temperatures $T_{\rm sub}$ and references for the optical constants. References are (1) \citet{dorschner1995} (2) \citet{jaeger2003} (3) \citet{laor1993} (4) \citet{zubko1996} (5) \citet{uspenskii2006}.}
  \begin{tabular}{cccc}
    \hline
    Dust species & $\rho_g$/${\rm g} \pcc$ & $T_{\rm sub}$/$\kel$ & $n$-$k$ \\
    \hline
    MgSiO$_3$ & $2.5$ & $1500$ & (1),(3) \\
    Mg$_{0.4}$Fe$_{0.6}$SiO$_3$ & $2.5$ & $1500$ & (1),(3) \\
    Mg$_{0.7}$SiO$_{2.7}$ & $2.5$ & $1500$ & (2),(3) \\
    Mg$_{2.4}$SiO$_{4.4}$ & $2.5$ & $1500$ & (2),(3) \\
    Am. carbon ACAR & $1.6$ & $2500$ & (4),(5) \\
    Am. carbon BE &  $1.6$ & $2500$ & (4),(5) \\
    \hline
  \end{tabular}
  \label{tab:casA:dustprop}
\end{table}

We initially assume an MRN size distribution \citep{mathis1977}, with $\amin = 0.005 \um$, $\amax = 0.25 \um$ and a power law size distribution with $\frac{dn}{da} \propto a^{-3.5}$, and calculate the SED per grain (averaged over size and temperature distributions) for each of the four components. As we do not consider dust self-absorption or other optically thick effects, the SEDs can be scaled to find the emission from an arbitrary number of grains, which can then be converted to a dust mass using the size distribution and grain density. Our observational data are the supernova dust fluxes reported by \citet{delooze2017} for $G = 0.6 G_0$, following the removal of line, synchrotron and ISM foreground emission, as listed in Table \ref{tab:casA:fluxes}. To fit the observed emission from Cas A, we run grids of models, with the number of dust grains in each component as the four free parameters, convolve the resulting SEDs with the filter profiles for each instrument and calculate the reduced $\chi^2$ values for the models. We do not include the IRAC $8 \um$ and WISE $12 \um$ points in our $\chi^2$ calculations, as at these wavelengths there is significant PAH emission from the ISM dust component in the THEMIS-based models of \citet{delooze2017} - as any real contribution from PAHs is highly uncertain, the reported SNR fluxes at these wavelengths cannot be used to constrain the dust emission.

\begin{table}
  \centering
  \caption{Cas A SNR dust fluxes and uncertainties from \citet{delooze2017}, for an ISM radiation field strength $G = 0.6 G_0$.}
  \begin{tabular}{ccc}
    \hline
    Waveband & $F_\nu / {\rm Jy}$ \\
    \hline
    IRAC $8 \um$ & $0.2 \pm 0.1$ \\
    WISE $12 \um$ & $3.4 \pm 0.3$ \\
    IRS $17 \um$ & $63.3 \pm 6.0$ \\
    WISE $22 \um$ & $202.0 \pm 19.3$ \\
    MIPS $24 \um$ & $153.4 \pm 15.0$ \\
    IRS $32 \um$ & $168.5 \pm 17.3$ \\
    PACS $70 \um$ & $149.5 \pm 20.1$ \\
    PACS $100 \um$ & $125.8 \pm 19.9$ \\
    PACS $160 \um$ & $69.9 \pm 12.0$ \\
    SPIRE $250 \um$ & $27.3 \pm 4.8$ \\
    SPIRE $350 \um$ & $10.9 \pm 1.9$ \\
    SPIRE $500 \um$ & $2.6 \pm 0.5$ \\
    SCUBA $850 \um$ & $0.4 \pm 0.1$ \\
    \hline
  \end{tabular}
  \label{tab:casA:fluxes}
\end{table}

\section{Results}

\begin{figure*}
  \centering
  \subfigure{\includegraphics[width=\columnwidth]{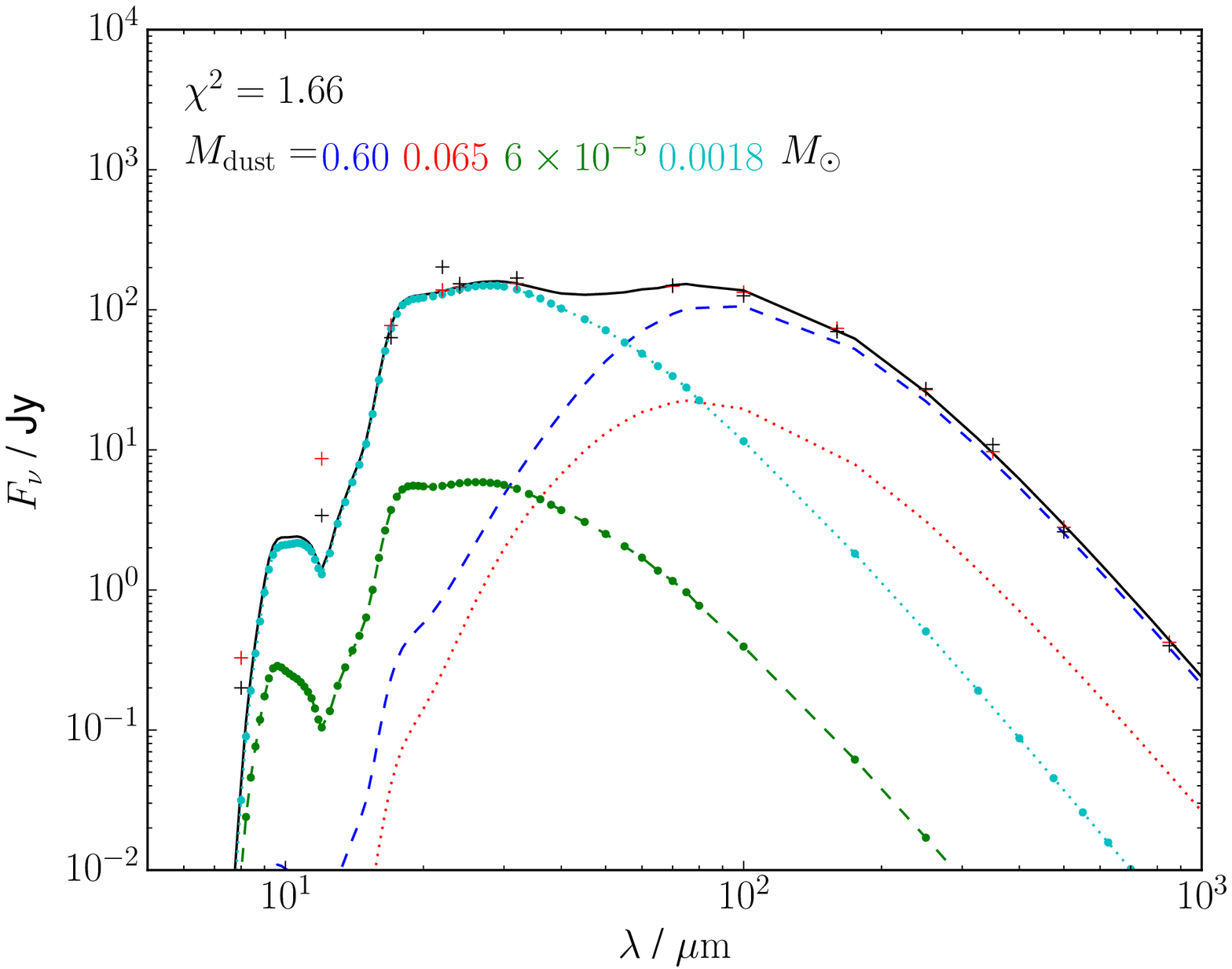}}\quad
  \subfigure{\includegraphics[width=\columnwidth]{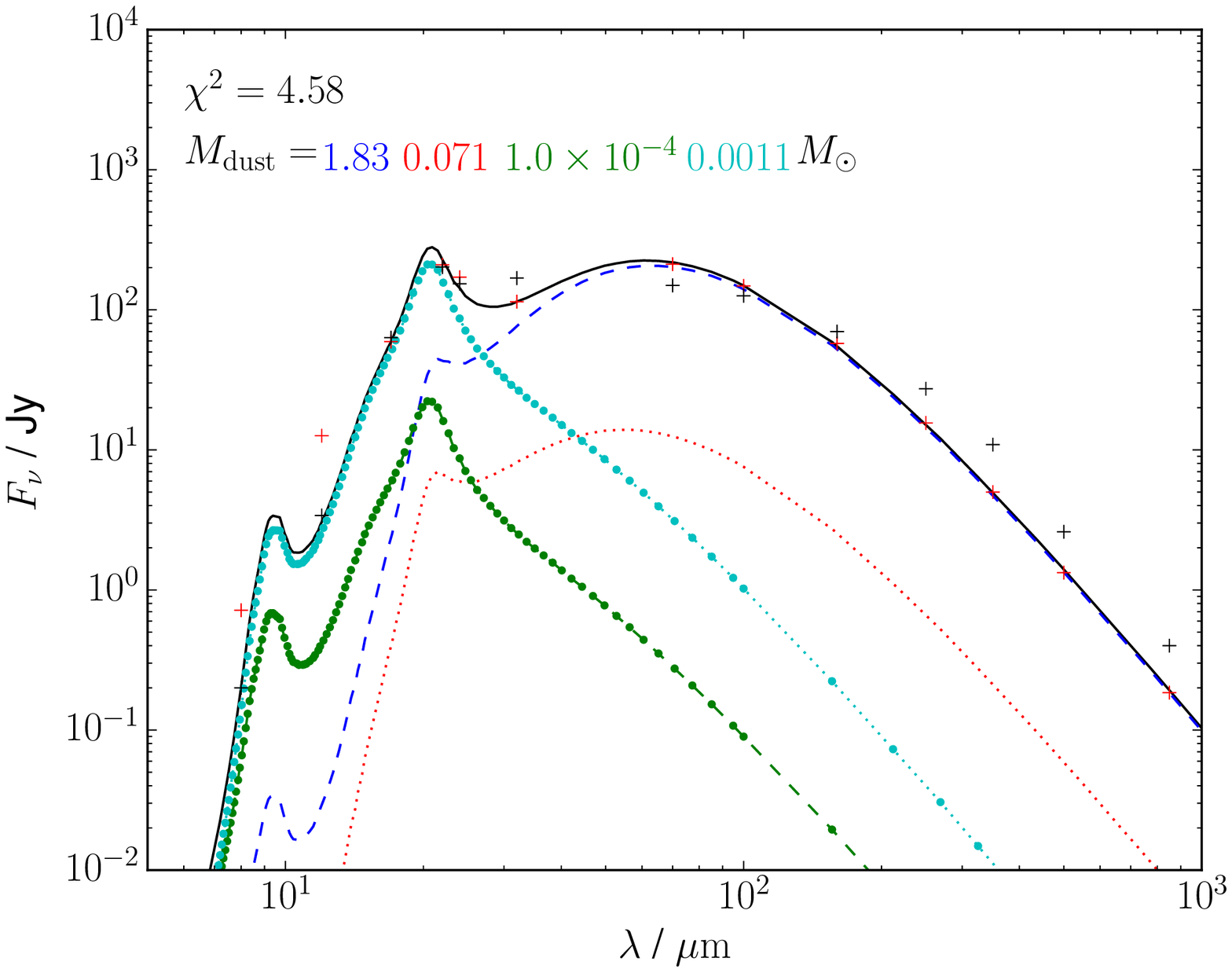}}\\
  \subfigure{\includegraphics[width=\columnwidth]{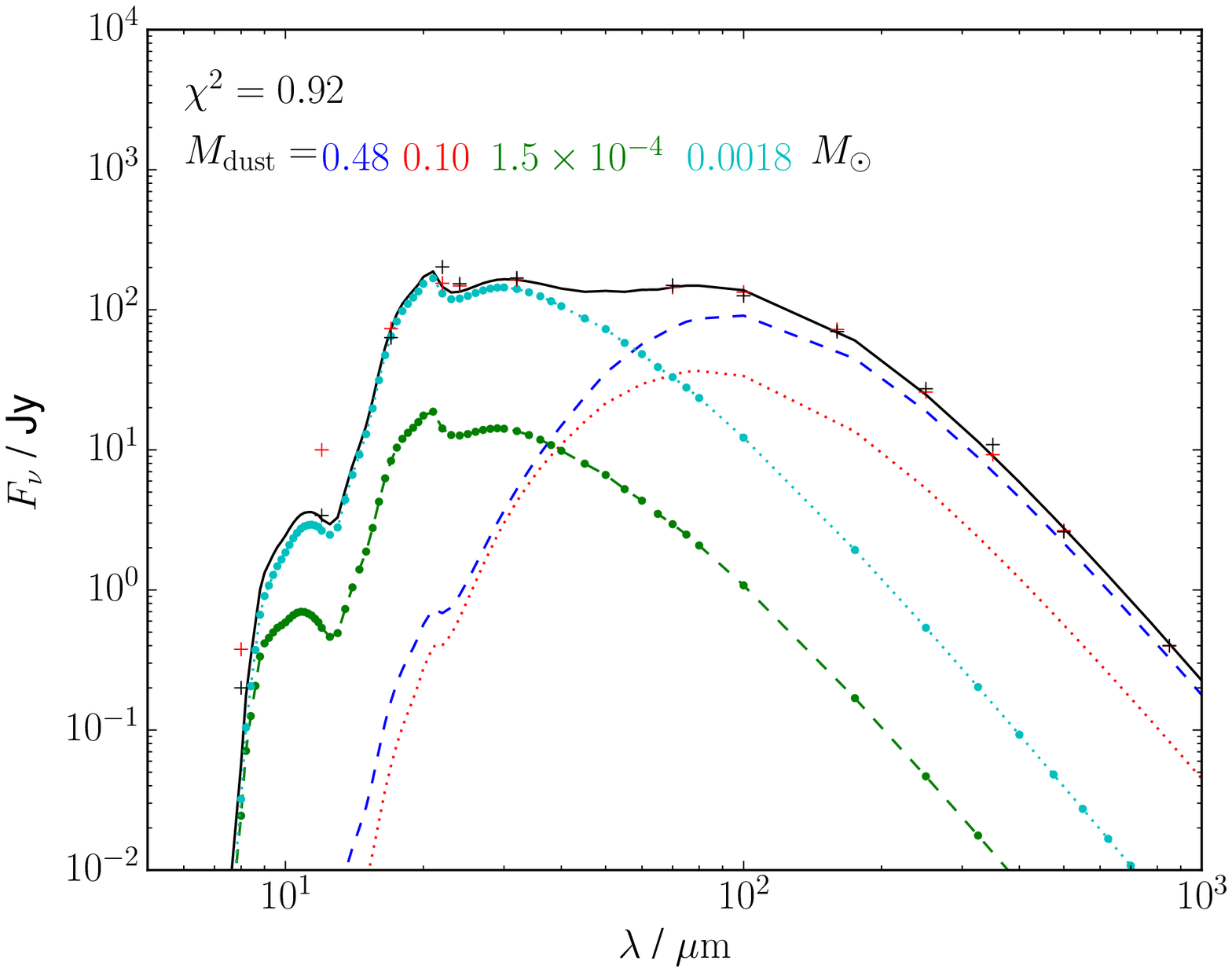}}\quad
  \subfigure{\includegraphics[width=\columnwidth]{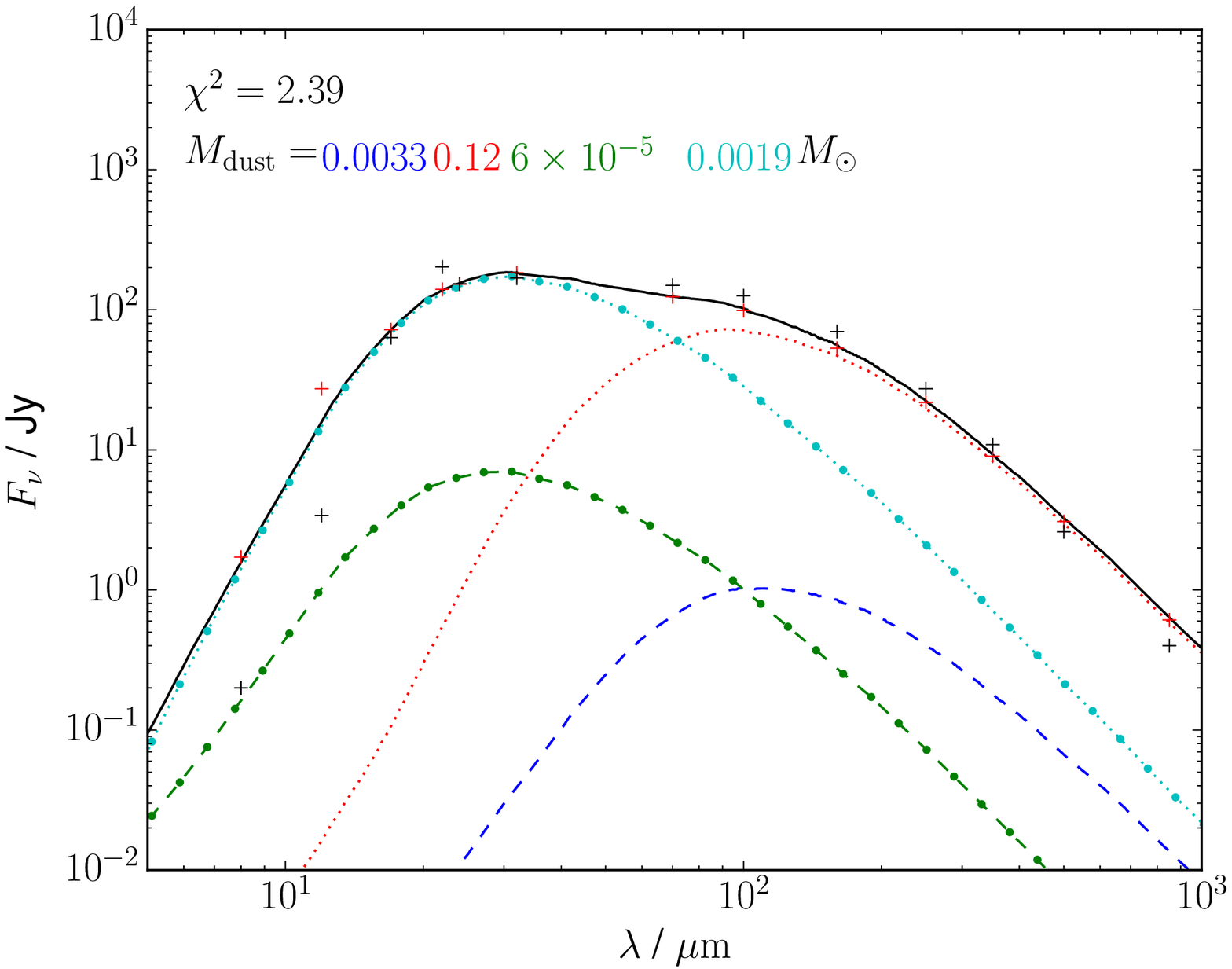}}
  \caption{$G=0.6 G_0$ Cas A dust fluxes from \citet{delooze2017} (black crosses), and the best-fit total dust SEDs (black solid line) and model fluxes (red crosses) for grains with an MRN size distribution. The SEDs from each dust component, as defined in Table \ref{tab:casA:gasprop}, are shown as blue dashed lines (preshock), red dotted lines (clumped), green circle-dashed lines (diffuse), and cyan circle-dotted lines (blastwave). The grain compositions are MgSiO$_3$ (upper left), Mg$_{0.7}$SiO$_{2.7}$ (upper right), Mg$_{0.4}$Fe$_{0.6}$SiO$_3$ (lower left) and ACAR (lower right).}
  \label{fig:casA:dustfit}
\end{figure*}

Figure \ref{fig:casA:dustfit} (upper left) shows our best-fit model SED for MgSiO$_3$ grains, with optical data from \citet{dorschner1995}. The reduced $\chi^2$ value, excluding the $8 \um$ and $12 \um$ points, is $1.66$. The model requires dust masses of $0.60$, $0.065$, $6 \times 10^{-5}$ and $0.0018 \msun$ for the preshock, clumped, diffuse and blastwave regions respectively, for a total dust mass of $0.67 \msun$, comparable to the value of $0.5 \pm 0.1 \msun$ found by \citet{delooze2017} for the same dust species. The \citet{delooze2017} `hot' dust component has a similar mass to the combined blastwave and diffuse components in our model, while their `warm' component has $\sim 10 \times$ less mass than our `clumped' dust. The average (mass-weighted) dust temperatures of the four components are $29$, $32$, $97$ and $94 \kel$ respectively. The cold, warm and hot dust temperatures found by \citet{delooze2017} are $30$, $79$ and $100 \kel$ - the preshock and clumped components produce the `cold' dust emission, while the diffuse and blastwave components reproduce the `hot' dust. The emission from the \citet{delooze2017} `warm' dust in our models originates from a combination of the smallest grain in the two `cold' components and the largest grains in the two `hot' components.  Our SED predicts more emission at $8$ and $12 \um$ than the reported SNR fluxes from \citet{delooze2017}, who did not include a high-temperature ($\sim 500 \kel$) dust component capable of fitting the short-wavelength emission - we note that our predicted SED in this region appears similar to the `$21 \um$ peak' IRS spectra of Cas A from \citet{rho2008}, reproducing the broad feature at $\sim 10 \um$.

Table \ref{tab:casA:dustmass} lists the best-fit dust masses and reduced $\chi^2$ values for each of the dust species listed in Table \ref{tab:casA:dustprop}. Table \ref{tab:casA:gasdust} lists the { dust-to-gas} mass ratios for each component, assuming the gas masses listed in Table \ref{tab:casA:gasprop}. The total dust masses for the four silicate species are similar, with the exception of Mg$_{0.7}$SiO$_{2.7}$, which required $\sim 3 \times$ as much mass. This species was also found to require a significantly larger dust mass than other silicates by \citet{delooze2017}, although their value of $21.4 \msun$ is much higher than ours, as they found a best-fit temperature of $21 \kel$ whereas even the largest dust grains are heated to $\sim 30 \kel$ in our model, thus requiring less mass to produce the same flux. The Mg$_{0.7}$SiO$_{2.7}$ SED is also a noticeably worse fit to the observations than the other models. Figure \ref{fig:casA:dustfit} (upper right) shows its best fit SED - the slope at long wavelengths is inconsistent with that observed, and the dip in flux beyond the $\sim 20 \um$ peak is too severe. Figure \ref{fig:casA:dustfit} (lower left) shows the best fit SED for Mg$_{0.4}$Fe$_{0.6}$SiO$_3$, the composition which gives the lowest $\chi^2$ value of those we consider.

The two carbon species considered show similar behaviour, with lower dust masses than the silicates, almost all of which is contained in the clumped component, rather than in the preshock component as with the silicates. Given that the preshock gas mass is several times larger than the clumped component, it seems unlikely that the dust mass is distributed in the opposite fashion. Our dust masses are somewhat lower than the results for carbon grains from \citet{delooze2017}, again due to our models having higher grain temperatures than their `cold' component, which contains most of the mass. Figure \ref{fig:casA:dustfit} (lower right) shows the best-fit SED for the ACAR grains. While the model provides an acceptable fit to the $> 20 \um$ data, the carbon dust SED at shorter wavelengths is drastically different to the IRS spectra from \citet{rho2008}, and the predicted flux at $12 \um$ is higher than that observed, before the subtraction of line, synchrotron and ISM contributions. Carbon grains also fail to reproduce the $21 \um$ feature.

\begin{table*}
  \centering
  \caption{Best-fit model dust masses { and (mass-weighted) average temperatures} for each component, total dust masses and reduced $\chi^2$ values for different grain species.}
  \begin{tabular}{c|ccccc|c}
    \hline
    & \multicolumn{5}{c|}{$M_{\rm dust}/\msun$ $(<T> / \kel)$} & \\
    Species & Preshock & Clumped & Diffuse & Blastwave & Total & $\chi^2$ \\
    \hline
    MgSiO$_3$ & $0.60$ $(29)$ & $0.065$ $(32)$ & $6 \times 10^{-5}$ $(97)$ & $0.0018$ $(94)$ & $0.67$ & $1.66$ \\
    Mg$_{0.4}$Fe$_{0.6}$SiO$_3$ & $0.48$ $(29)$ & $0.10$ $(33)$ & $1.5 \times 10^{-4}$ $(97)$ & $0.0018$ $(95)$ & $0.58$ & $0.92$ \\
    Mg$_{0.7}$SiO$_{2.7}$ & $1.83$ $(40)$ & $0.071$ $(45)$ & $1.0 \times 10^{-4}$ $(110)$ & $0.0011$ $(108)$ & $1.90$ & $4.58$ \\
    Mg$_{2.4}$SiO$_{4.4}$ & $0.56$ $(29)$ & $0.11$ $(33)$ & $6 \times 10^{-5}$ $(98)$ & $0.0018$ $(96)$ & $0.67$ & $1.32$ \\
    Am. carbon ACAR & $0.0033$ $(28)$ & $0.12$ $(33)$ & $6 \times 10^{-5}$ $(110)$ & $0.0019$ $(105)$ & $0.13$ & $2.39$ \\
    Am. carbon BE & $0.002$ $(30)$ & $0.19$ $(34)$ & $0.0$ $(114)$ & $0.0019$ $(108)$ & $0.19$ & $2.07$ \\
    \hline
  \end{tabular}
  \label{tab:casA:dustmass}
\end{table*}

\begin{table*}
  \centering
  \caption{Best-fit model { dust-to-gas} mass ratios for each component for different grain species.}
  \begin{tabular}{c|cccc}
    \hline
    Species & Preshock & Clumped & Diffuse & Blastwave \\
    \hline
    MgSiO$_3$ & $0.20$ & $0.11$ & $3.6 \times 10^{-5}$ & $2.2 \times 10^{-4}$ \\
    Mg$_{0.4}$Fe$_{0.6}$SiO$_3$ & $0.16$ & $0.17$ & $8.9 \times 10^{-5}$ & $2.2 \times 10^{-4}$ \\
    Mg$_{0.7}$SiO$_{2.7}$ & $0.63$ & $0.12$ & $6.0 \times 10^{-5}$ & $1.3 \times 10^{-4}$ \\
    Mg$_{2.4}$SiO$_{4.4}$ & $0.19$ & $0.19$ & $3.6 \times 10^{-5}$ & $2.2 \times 10^{-4}$ \\
    Am. carbon ACAR & $0.0011$ & $0.20$ & $3.6 \times 10^{-5}$ & $2.3 \times 10^{-4}$ \\
    Am. carbon BE & $6.7 \times 10^{-4}$ & $0.32$ & $0.0$ & $2.3 \times 10^{-4}$ \\
    \hline
  \end{tabular}
  \label{tab:casA:gasdust}
\end{table*}

\begin{figure}
  \centering
  \includegraphics[width=\columnwidth]{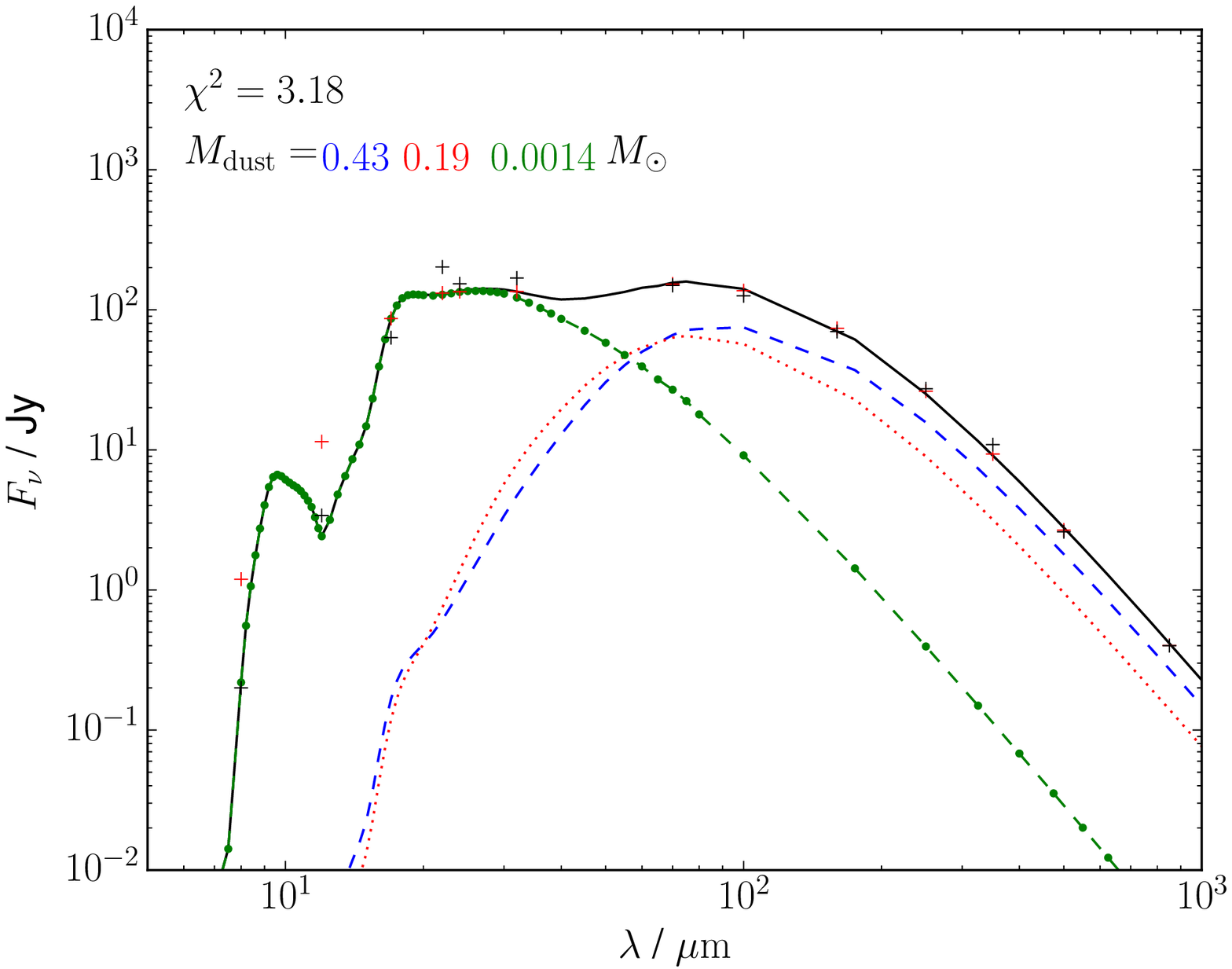}
  \caption{$G=0.6 G_0$ Cas A dust fluxes from \citet{delooze2017} (black crosses), and the best-fit total dust SEDs (black solid line) and model fluxes (red crosses) for MgSiO$_3$ grains with an MRN size distribution, using only three ejecta components. The SEDs from each dust component as defined in Table \ref{tab:casA:gasprop} are shown as blue dashed lines (preshock), red dotted lines (clumped) and green circle-dashed lines (diffuse).}
  \label{fig:casA:dustaltfit}
\end{figure}

The three grain species that produce good fits to the observed fluxes predict similar dust masses for each gas component, with the majority in the preshock region and a smaller amount in the clumps. The fitted dust masses in the two X-ray emitting components are insignificant in terms of the total dust mass but are essential for reproducing the short wavelength data. The best-fit models all predict most of this high-temperature dust to be in the blast wave component. However, given the similarity of the blast wave and diffuse dust SEDs there is a large amount of degeneracy between them. Figure \ref{fig:casA:dustaltfit} shows an SED fit for only the three ejecta components. While the $\chi^2$ is worse than for models with all four components, the fit is still acceptable, with the diffuse dust mass comparable to the values found for the blast wave component in Table \ref{tab:casA:dustmass}. The total dust mass is close to the four-component fit value, although the fraction of dust in the clumps is larger.

{ The clumped SED, although similar to the preshock component, has additional emission at wavelengths $\lesssim 100 \um$, which can be seen in Figure \ref{fig:casA:dustfit}, where the clumped and preshock SEDs are comparable at these wavelengths despite the preshock SED being dominant for $\lambda > 100 \um$. As the diffuse/blastwave components also emit strongly in this region, the amount of dust in the clumped component, and its contribution to the longer wavelength SED, is constrained. The remainder of the $> 100 \um$ emission must come from the preshock component, so the degeneracy is broken and the two components can be distinguished.}

\subsection{Implications of model assumptions}

While the temperatures and densities of the matter in the two X-ray emitting components have been derived from observations \citep{willingale2003}, the two cooler components, which contain most of the dust mass, have more uncertain properties. The gas temperatures are fairly well constrained, while \citet{smith2009} found an upper limit of $\nel \lesssim 100 \pcc$ for the preshock ejecta, but the reported densities for the clumped material range up to $10^5 \pcc$, far higher than our model value of $480 \pcc$. Both dust components are primarily heated by the synchrotron radiation field, which we obtained by interpolating between radio and X-ray measurements and assuming a mean distance of $1 \pc$ from the synchrotron radiation source in the blastwave - the actual radiation field could differ in both strength and spectral shape. We have also assumed an MRN size distribution for all dust components, which may not be justified given the expected growth of grains in the preshock component and the processing of grains in the reverse shock. In the following, we assess the impact of each of these parameters on the resulting dust masses. We use MgSiO$_3$ grains unless otherwise stated.

\subsubsection{Gas density}

At the preshock ejecta temperature of $100 \kel$ \citep{raymond2018}, the effects of collisional heating on the dust emission are negligible - even increasing the electron and ion densities to $10^4 \pcc$, we find no noticeable difference in the model SED. For the clumped component, a density of $10^4 \pcc$ causes electron collisional heating to dominate the heating rate, and the dust emission becomes stronger and shifted towards shorter wavelengths. Repeating our fitting procedure, we find that the dust mass in the clumped component is reduced by a factor of a few, while the preshock component mass increases slightly so that the total dust mass is barely affected. The mass in the diffuse component is reduced to negligible levels in this case.

\subsubsection{Radiation field}

\begin{figure*}
  \centering
  \subfigure{\includegraphics[width=\columnwidth]{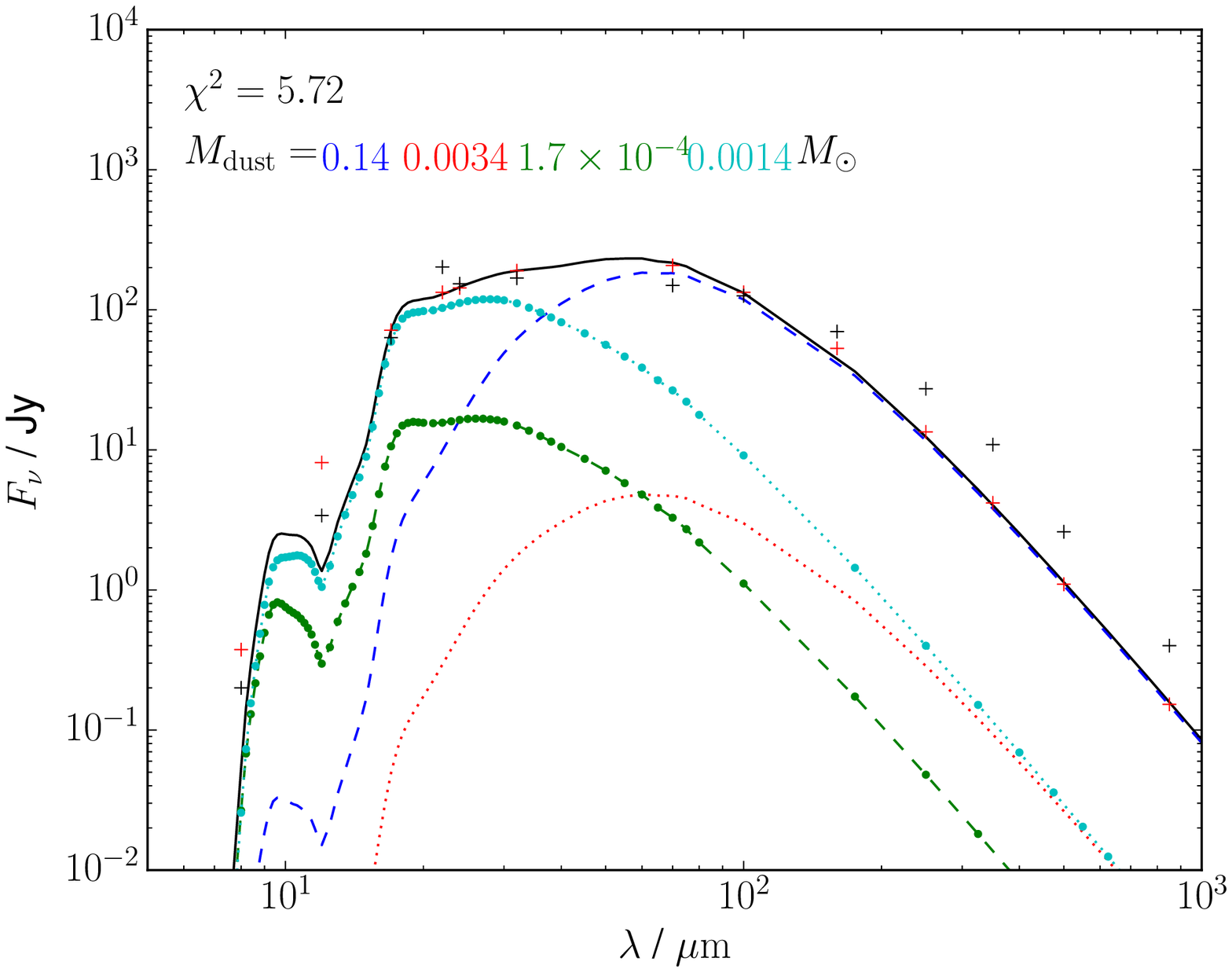}}\quad
  \subfigure{\includegraphics[width=\columnwidth]{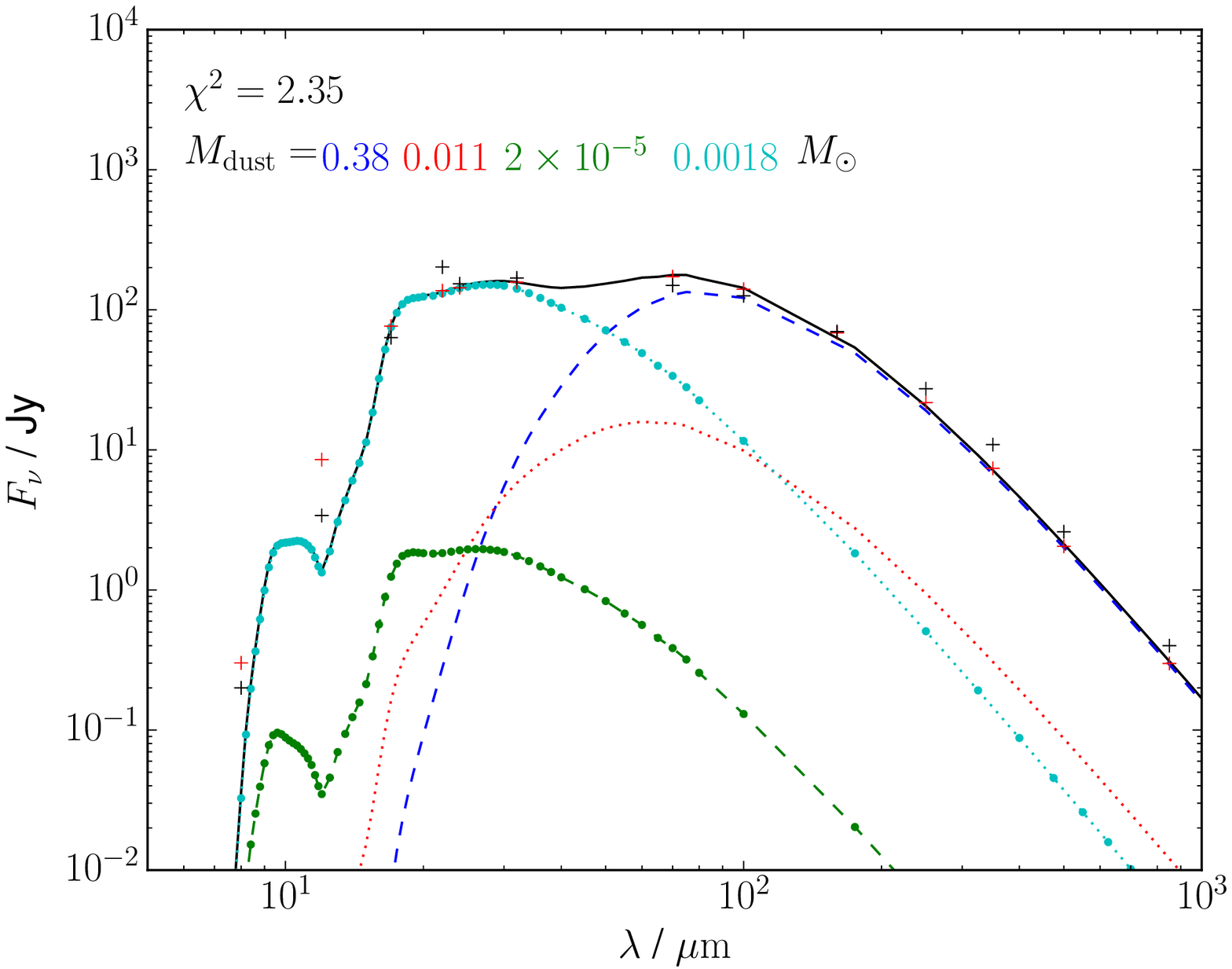}}
  \caption{$G=0.6 G_0$ Cas A dust fluxes from \citet{delooze2017} (black crosses), and the best-fit total dust SEDs (black solid line) and model fluxes (red crosses) for MgSiO$_3$ grains with the synchrotron radiation field strength increased by a factor of $10$. The SEDs from each dust component as defined in Table \ref{tab:casA:gasprop} are shown as blue dashed lines (preshock), red dotted lines (clumped), green circle-dashed lines (diffuse) and cyan circle-dotted lines (blastwave). In the left panel, all components have MRN size distributions; in the right panel, the preshock component values of $\amin/\amax$ have been changed to $0.1/1.0 \um$.}
  \label{fig:casA:dustfitrad10}
\end{figure*}

To investigate the sensitivity of our results to the synchrotron radiation field, probably the least certain of our model inputs, we increased its strength by a factor of $10$ for all components. Figure \ref{fig:casA:dustfitrad10} (left) shows the best fit SED for MgSiO$_3$ grains. The required dust masses in the preshock and clumped components are reduced to $\sim 0.1 \msun$, while the diffuse and blast wave components, which are heated by particle collisions, are less affected. The model is a poor fit, in particular to the long-wavelength data, where it predicts significantly less flux than observed. We find that this can be remedied by changing the size distribution in the unshocked component to grains only between radii of $0.1-1.0 \um$. This is shown in Figure \ref{fig:casA:dustfitrad10} (right) - the dust mass is still lower than models with the original choice of radiation field, but only by about a third, and the proportion in the preshock component is even higher. Increasing the radiation field by smaller factors also requires an increase in the number of large grains in the preshock component in order to reproduce the long-wavelength flux, with the total dust mass being mostly unchanged even for a $5 \times$ increase in the radiation field.

\subsubsection{Grain size distribution}

Altering the grain size power law exponent to $2$ results in an increased total dust mass of $0.76 \msun$, with $0.084 \msun$ in the preshock component and $0.67 \msun$ in the clumped component, although for these models these two components are almost entirely degenerate. The reduced $\chi^2$ values is $1.24$ - all models in the following discussion have $\chi^2 \lesssim 2$. In this case the dust mass in the diffuse component ($0.003 \msun$ dominates over the blastwave component ($<10^{-4} \msun$), although again with some degeneracy between the two. For a power law exponent of $4$, results are similar to the MRN case, although the total dust mass is slightly lower ($0.54 \msun$). Changing $\amin$ and $\amax$ between $0.001-0.01$ and $0.05-1.0 \um$ respectively, acceptable fits can still be obtained, with the total dust mass staying within the range $0.4-0.8 \msun$ in all cases.

\section{Discussion}

\subsection{Comparison with previous results}

Our total dust masses are comparable to those found by \citet{delooze2017} - given that we use their best-fit SNR dust fluxes as input, this is unsurprising. Although not directly equivalent, their `cold' dust component corresponds to the flux produced by our preshock and clumped components, while their `warm' and `hot' dust corresponds to the two X-ray emitting components. The required dust masses are greater than several previous studies of the infrared emission found for Cas A \citep{rho2008,barlow2010,arendt2014}, generally $\lesssim 0.1 \msun$ - as discussed by \citet{delooze2017}, this is due to the inclusion of better defined long wavelength data, allowing the cold dust contribution to be better determined. Modelling of Cas A's integrated emission line profiles by \citet{bevan2017} found a dust mass of $\sim 1 \msun$, comparable to our values.

As well as the total dust mass, our preferred dust compositions are also similar to those found by \citet{delooze2017}, with various magnesium-containing silicates (with the exception of Mg$_{0.7}$SiO$_{2.7}$) all proving acceptable. While \citet{delooze2017} suggested some carbon dust could also be present, we find that even a small ($\sim 10^{-3} \msun$) mass of carbon dust in the X-ray emitting gas predicts near-infrared fluxes in conflict with observations, although larger carbon dust masses could be present in the cooler regions. \citet{bevan2017} found that mixtures of carbon and silicate grains can explain line-profile asymmetries in Cas A, but increasing the proportion of silicates required increasingly large dust masses, up to unreasonably high values ($6.5 \msun$ for $90 \%$ silicates). Since the unshocked ejecta would be responsible for the majority of the line asymmetry, this is not necessarily in conflict with our results, although the lack of carbon in the shocked ejecta would require explanation if it makes up a significant fraction of the unshocked ejecta. Multiple dust species have been used by \citet{rho2008} and \citet{arendt2014} to fit the mid-infrared spectrum, which we discuss further in the next section, but some form of magnesium silicate was required by both those studies.

Theoretical studies of the formation of dust in the specific case of Cas A have led to differing results. \citet{nozawa2010} predicted a total dust mass of $0.167 \msun$ with roughly equal quantities of carbon and magnesium silicates, and smaller contributions from other species. \citet{bocchio2016} predicted an even higher dust mass of $0.92 \msun$, about half of which is in silicates and a smaller fraction in carbon, while \citet{biscaro2016} predicted only $\sim 10^{-2} \msun$ with the main dust component being Al$_2$O$_3$. Of these studies only that of \citet{bocchio2016} is consistent with our results - the others predict dust masses significantly smaller than our required values, and of different composition. While we did not investigate Al$_2$O$_3$, \citet{delooze2017} found that the required mass of aluminium for this composition significantly exceeded that expected from calculations of the nucleosynthetic yields by \citet{woosley1995}.

\subsection{Heating mechanisms}

As mentioned previously, the preshock and clumped dust components are heated primarily by the synchrotron radiation field, while the diffuse and blastwave components are heated by particle collisions. However, the additional collisional heating due to the higher temperature in the clumped component causes a subtle but important difference in the dust SED, compared to the preshock dust, which results in our best-fit models having the majority of the dust mass in the preshock ejecta. This shows the importance of treating both processes simultaneously. \citet{bocchio2013} have previously investigated the effects of the addition of electron collisional heating in the ISM, but calculations of dust heating in SNRs have generally considered only radiative (e.g. \citealt{temim2013,owen2015}) or collisional \citep{dwek1987,dwek2008} effects.

We also find that in the diffuse and blastwave components, the heating rate from collisions with nuclei is comparable to that from electrons ($\sim 0.3-0.5$ of the total heating rate), with the difference between $\tion$ and $\tel$ enough to counteract the higher mass of the ions (and subsequent reduction in velocities and collision rates). As this non-equilibrium between ions and electrons is expected to be a general feature of shocks in SNRs \citep{raymond2018b}, it is not necessarily true that electrons will be the dominant source of dust heating even in situations where collisional heating can be expected to dominate.

\subsection{The mid-infrared spectrum}

\citet{rho2008} and \citet{arendt2014} both used {\it Spitzer} IRS $5-30 \um$ spectra of Cas A, in addition to longer-wavelength photometric data extending to $160 \um$, to investigate the dust composition in different regions of the SNR. \citet{rho2008} categorized the spectra based on the strength of the $21 \um$ peak compared to the continuum, whereas \citet{arendt2014} defined various regions of the remnant characterised by particular dominant emission features (line or continuum), and treated the dust emission from these regions as separate populations. \citet{rho2008} found $0.02-0.05 \msun$ of dust to be responsible, with the dominant materials being carbon and simple iron and silicate species (e.g. FeO, SiO$_2$), although also requiring some form of magnesium silicate and Al$_2$O$_3$. \citet{arendt2014} found $\sim 0.04 \msun$ of dust to be needed, which could mostly be fit by a combination of magnesium silicates and some additional featureless dust component (carbon, Al$_2$O$_3$ or Fe/FeS/Fe$_3$O$_4$), plus an additional featureless dust component of $\lesssim 0.1 \msun$ associated with the [Si II] emission, which they could not constrain the composition of but associated with the unshocked ejecta.

Both studies found dust masses lower than our values by an order of magnitude - this is again due to the additional long wavelength fluxes we use, where the majority of the dust mass emits. \citet{arendt2014} did include Herschel PACS fluxes to constrain the dust masses, in particular the [Si II] component, from which they determined an upper limit of $0.1 \msun$ of unshocked dust, well below our values of $\sim 0.5 \msun$. However, in determining the ISM dust contribution at $160 \um$ they assumed a scaled version of the SPIRE $250 \um$ map to represent the ISM dust emission, whereas \citet{delooze2017} find a substantial SNR contribution to the $250 \um$ flux. Their SNR $160 \um$ flux, and the derived dust mass, are therefore lower than the values from \citet{delooze2017} which took into account ISM and SNR dust emission simultaneously.

Although there are variations from position to position, the main features of the mid-IR spectra are two emission peaks of varying strength at $\sim 10 \um$ and $21 \um$, and an underlying continuum which rises to $21 \um$ and stays roughly constant or falls to longer wavelengths. While \citet{rho2008} and \citet{arendt2014} used various combinations of dust species and temperature components to fit this behaviour, our model SEDs approximately reproduce this naturally using single silicate species - the two peaks and the rising part of the continuum are produced by grains in the high temperature X-ray emitting parts of the ejecta, while dust in the cooler regions produces an increasing fraction of the flux towards longer wavelengths, preventing the decline found with single-temperature $\sim 100 \kel$ SEDs. Some IRS spectra (`featureless' in \citet{rho2008}, `[Ne II]' in \citet{arendt2014}) required an additional dust component at $\sim 10 \kel$ to explain featureless emission, possibly carbon or Al$_2$O$_3$, but we find that the observations are consistent with magnesium silicate grains making up the majority of the dust present in the SNR.

\begin{figure*}
  \centering
  \includegraphics[width=2\columnwidth]{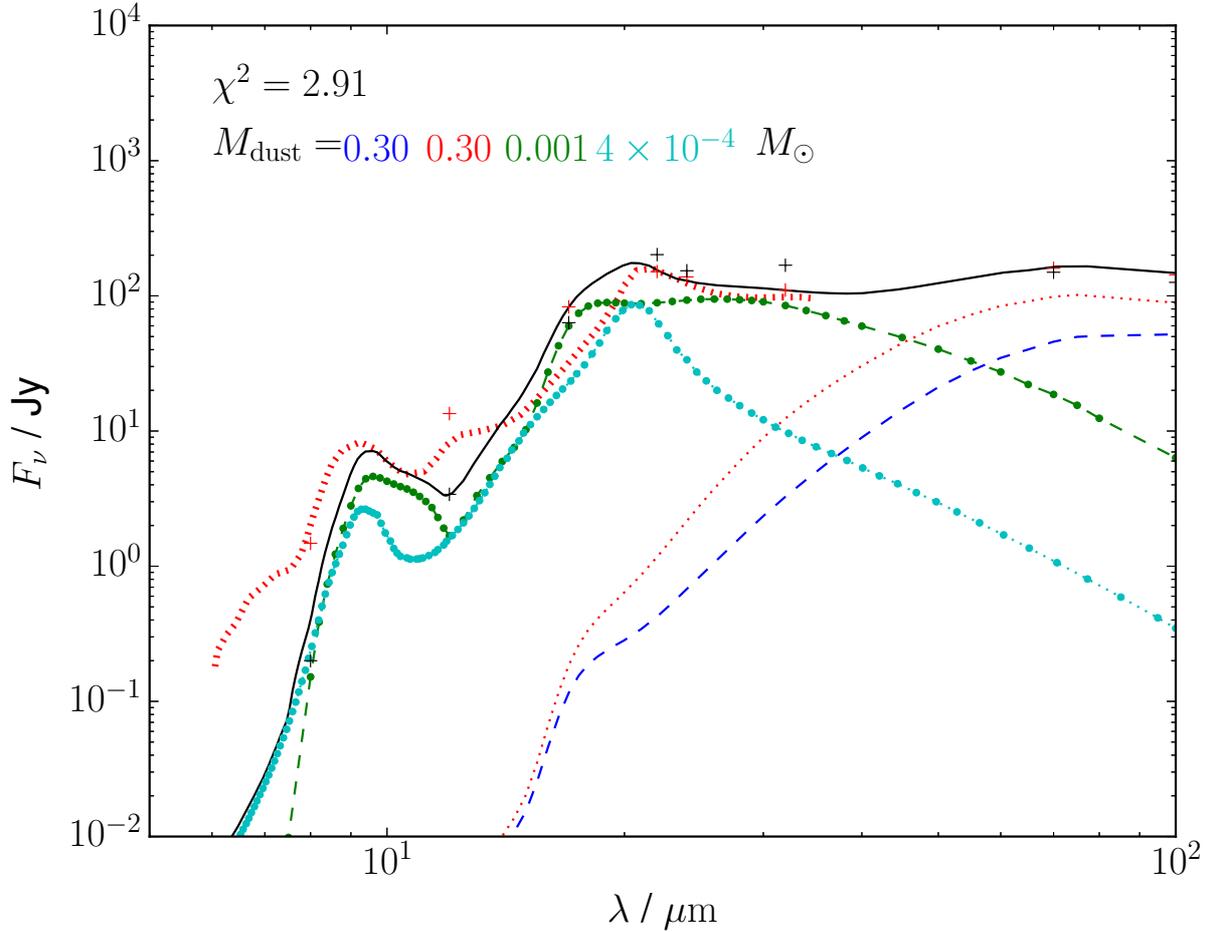}
  \caption{$G=0.6 G_0$ Cas A dust fluxes from \citet{delooze2017} (black crosses), and the best-fit total dust SEDs (black solid line) and model fluxes (red crosses) for MgSiO$_3$ and Mg$_{0.7}$SiO$_{2.7}$ grains with an MRN size distribution. The SEDs from dust components as defined in Table \ref{tab:casA:gasprop} are shown as blue dashed lines (preshock), red dotted lines (clumped), green circle-dashed lines (diffuse MgSiO$_3$) and cyan circle-dotted lines (diffuse Mg$_{0.7}$SiO$_{2.7}$). The `[Ar II] region' dust spectrum from \citet{arendt2014} is shown as a stippled red curve.}
  \label{fig:21um}
\end{figure*}

Our best-fit magnesium silicate models all underpredict the flux at $21 \um$, with the exception of Mg$_{0.7}$SiO$_{2.7}$, which otherwise provides a poor fit. The shape of our predicted SED in this region also clearly differs from many of the individual IRS spectra (the `$21 \um$ peak dust' of \citealt{rho2008}), suggesting that some additional component may be contributing at this wavelength. \citet{arendt2014} attribute this component to Mg$_{0.7}$SiO$_{2.7}$, while \citet{rho2008} adopt a combination of FeO and SiO$_2$. By including an additional dust component from one of these species, assuming FeO and SiO$_2$ can be treated as silicates and using the diffuse X-ray emitting gas properties, we find that this $21 \um$ excess can be reproduced with the addition of $\sim 5 \times 10^{-4} \msun$ of dust. FeO and Mg$_{0.7}$SiO$_{2.7}$ both produce similar SEDs, while SiO$_2$ produces an additional sharp peak at $\sim 12 \um$, similar to that seen in some of the IRS spectra (e.g. Figure 3 of \citet{rho2008}) and not produced by other silicate species. Figure ~\ref{fig:21um} shows the best-fit dust SED for MgSiO$_3$ as before, except with the blastwave component replaced by Mg$_{0.7}$SiO$_{2.7}$ using diffuse component properties, and the [Ar II] dust spectrum from \citet{arendt2014} (their Figure 3), scaled to the intensity of the total Cas A IRS spectrum at $21 \um$. We note that the model SED is a reasonable fit to the observed spectrum, in particular the feature at $\sim 10 \um$, despite only being fit to the flux points at $> 17 \um$ and not to the spectral data at all.

\subsection{Dust-to-gas mass ratios}

For the three dust species considered that produce acceptable fits to the observed SED (MgSiO$_3$, Mg$_{0.4}$Fe$_{0.6}$SiO$_3$ and Mg$_{2.4}$SiO$_{4.4}$), we find similar dust masses in each of the four components - $\sim 0.5-0.6 \msun$ in the preshock ejecta, $\sim 0.1 \msun$ in the clumps, and $10^{-4}/2 \times 10^{-3} \msun$ in the X-ray emitting reverse shock/blast wave gas. Taking the gas masses of these components as $3.0$ \citep{arias2018}, $0.59$ (Appendix \ref{sec:clump}), and $1.68$ and $8.32 \msun$ (both from \citealt{willingale2003}) respectively, this gives { dust-to-gas mass ratios of $0.20$, $0.17$, $6.0 \times 10^{-5}$ and $2.4 \times 10^{-4}$ (Table \ref{tab:casA:gasdust}), compared to a typical Galactic ISM value of $\sim 0.0067$ \citet{draine2011}.} The gas mass of $3 \msun$ obtained by \citet{arias2018} for the preshock component was noted by those authors to be higher than expected by many models of the emission from the shocked gas for the Cas A SNR, and could be lower if the unshocked ejecta gas has a lower temperature than assumed, or has a clumpy structure. \citet{raymond2018} found a temperature of $\sim 100 \kel$ for this component, the same value used by \citet{arias2018} to derive the $3 \msun$ gas mass, but the level of clumping is still uncertain. The { dust-to-gas ratio for this component may therefore be a lower limit} (assuming our dust mass is accurate), while for the other components the gas masses are better defined. We note that assuming a lower total (preshock + clumped + diffuse) ejecta gas mass (e.g. $2 \msun$; \citealt{laming2003}) would result in extremely { high ($\gtrsim 1$) dust-to-gas} ratios in either the preshock or clumped components, given the relatively well-constrained diffuse mass and the need for $\sim 0.5 \msun$ of dust in one of the other two components to reproduce the far-IR flux.

The ejecta { dust-to-gas ratios are $\sim 20 \times$ higher} than in the ISM, implying that a significant fraction of the metals are condensed into dust grains ($0.17$, assuming all the ejecta mass is condensible material). \citet{raymond2018} estimated a similar effiency of $\sim 0.1$ from the gas and dust masses from \citet{arias2018} and \citet{delooze2017} respectively, combined with their measurement of the preshock gas temperature. \citet{owen2015} found { lower ratios ($0.026-0.038$)} for the Crab Nebula, which has no reverse shock processing the ejecta - however, the Crab Nebula contains significant quantities of hydrogen and helium, unlike Cas A, so the fraction of condensible material locked up in dust grains is higher than the value inferred from this ratio. \citet{nozawa2010} found a condensation efficiency of $0.13$ in their model of dust formation in Cas A, although they predicted a lower dust mass than our models require, and also predicted that the main dust component is carbon, which we rule out. The { lower-than-ISM dust-to-gas mass ratios} in the two diffuse shocked components are consistent with significant dust destruction by both the forward and reverse shocks. Even if the blastwave dust mass is attributed instead to the reverse shock (which is possible, due to the degeneracy between the two components), the { dust-to-gas ratio would be $0.0013$}, implying that $< 1\%$ of the original dust mass has survived, assuming the original ratio is that of the preshock component. We note that dust condensation efficiencies are expected to be lower in the outer ejecta than the centre, which may suggest a { lower initial dust-to-gas ratio} for the material which has been processed by the reverse shock at present, and a correspondingly higher survival fraction. The surviving dust mass in the Cas A model of \citet{bocchio2016} is $\sim 1 \%$ of the initial mass.

Conversely, we find that the { dust-to-gas ratio} in the clumps is very similar to that in the the preshock ejecta, suggesting that dust within the clumps has been protected from destruction by the reverse shock. \citet{biscaro2016} predicted a surviving dust mass fraction of $6-11 \%$ for clumps in the Cas A SNR, while \citet{micelotta2016} predicted a surviving fraction of $12 \%$ for silicate grains, lower than our implied value even for conservative estimates of the unshocked ejecta gas mass - for a preshock gas mass of $1 \msun$ the surviving fraction is $28 \%$, rising to $84 \%$ for the $3 \msun$ value given by \citet{arias2018}. Dust masses from an increased radiation field model (Figure \ref{fig:casA:dustfitrad10}) suggest a surviving fraction of $15 \%$, closer to the theoretical estimates, although this would require all the dust to be located at an average distance of $\sim 0.3 \pc$ from the synchrotron radiation source. \citet{biscaro2016} and \citet{micelotta2016} include sputtering of dust grains expelled from the clumps into the inter-clump medium - accounting only for sputtering within the clumps before injection into the inter-clump medium, the value from \citet{biscaro2016} is $28-58 \%$, which is consistent with our results if the unshocked ejecta gas mass was $\sim 1-2 \msun$. The dust production efficiency of SNRs, and their overall contribution to the dust budget in the ISM, can therefore be strongly affected by the degree of clumping and the detailed, multi-dimensional hydrodynamical evolution of the ejecta as it passes through the reverse shock.

\section{Conclusions}

We have modelled the emission from dust grains subjected to the physical conditions present in the Cas A SNR, accounting for radiative and collisional heating mechanisms using a new dust emission code {\sc dinamo}, and used the SNR dust fluxes describing the IR SED from $17-850 \um$ from \citet{delooze2017} to constrain both the mass of dust present, and its distribution between the various components of the remnant (unshocked ejecta, shocked ejecta clumps, diffuse shocked ejecta and material swept up by the blast wave). We find dust masses of $\sim 0.6-0.7 \msun$ depending on the silicate composition, with the majority being located in the unshocked ejecta ($\sim 90 \%$) and postshock clumps ($\sim 10 \%$), and only a small fraction present in the hot X-ray emitting gas. The { dust-to-gas ratio in the shocked clumpy ejecta is $\sim 0.17$, similar to that in the unshocked region ($\sim 0.2$), while in the diffuse components it is significantly lower ($\lesssim 0.001$).} This is consistent with dust grains being efficiently sputtered at high temperatures, whereas in the ejecta clumps which have passed through the reverse shock, the dust is more resilient to destruction. Magnesium silicate grains, with possible iron inclusions, are found to reproduce almost all of the observed Cas A dust spectrum, with a relatively minor amount of another species (FeO, SiO$_2$ or Mg$_{0.7}$SiO$_{2.7}$) required to reproduce the $21 \um$ emission peak. While carbon grains may be present, they cannot make up a large fraction of the dust mass in the X-ray emitting gas ($< 25 \%$) without predicting NIR fluxes in excess of those observed. If the mass fractions do not vary significantly between the shocked and unshocked ejecta, carbon dust can be ruled out as a major constituent of the ejecta dust. The unshocked and clumped ejecta dust, making up the majority of the mass, is heated mostly by the remnant's synchrotron radiation field, while the diffuse and blastwave dust, which dominates the total SED luminosity, is heated by collisions with electrons and nuclei. The total dust mass in Cas A is consistent with CCSNe being significant contributors to the dust in high-redshift galaxies, particularly if much of it is present in clumps which survive the passage of the reverse shock without disruption.

\section*{Acknowledgements}

FDP is supported by the Science and Technology Facilities Council. MJB acknowledges support from the European Research Council grant SNDUST ERC-2015-AdG-694520. IDL gratefully acknowledges the support of the Research Foundation -- Flanders (FWO).

%%%%%%%%%%%%%%%%%%%%%%%%%%%%%%%%%%%%%%%%%%%%%%%%%%



\bibliographystyle{mnras}
\bibliography{casA}

%%%%%%%%%%%%%%%%%%%%%%%%%%%%%%%%%%%%%%%%%%%%%%%%%%

%%%%%%%%%%%%%%%%% APPENDICES %%%%%%%%%%%%%%%%%%%%%

\appendix

\section{Dust heating and cooling rates}
\label{sec:rates}

\subsection{Radiative heating}

A dust grain in enthalpy bin $i$, with enthalpy $H_i$, absorbing a photon of energy $\hv$ will increase its enthalpy to $H_i + \hv$, which may move it into a higher enthalpy bin. The transition rate between bins $i$ and $f$ due to radiative heating is given by
\begin{equation}
  A_{fi} = 4 \pi^2 a^2 Q_{\lambda_{fi}} J_{\lambda_{fi}} \frac{hc \Delta H_f}{\left( H_f - H_i \right)^3}
\end{equation}
for $f > i$, where $a$ is the grain radius, $\lambda_{fi} = \frac{hc}{H_f - H_i}$ is the wavelength of a photon with energy corresponding to the difference between enthalpy bins, $Q_{\lambda_{fi}}$ and $J_{\lambda_{fi}}$ are the absorption efficiency and radiation field strength at wavelength $\lambda_{fi}$ and $\Delta H_f$ is the width of enthalpy bin $f$ \citep{camps2015}. Photons energetic enough to heat the grain beyond the highest enthalpy bin are included in the rate to bin $N$, giving an additional term
\begin{equation}
  A'_{Ni} = 4 \pi^2 a^2 \int_0^{\lambda_{\rm max}} \lambda Q_{\lambda} J_{\lambda} d \lambda / hc
\end{equation}
where $\lambda_{\rm max} = \frac{hc}{H_{\rm max} - H_i}$ is the wavelength of the least energetic photon capable of heating a grain beyond the maximum temperature. Photons not energetic enough to heat a grain out of the enthalpy bin contribute to a continuous heating rate
\begin{equation}
  \frac{dH_{\rm heat}}{dt} = 4 \pi^2 a^2 \int^{\infty}_{\lambda_0} Q_{\lambda} J_{\lambda} d \lambda
\end{equation}
where $\lambda_0 = \frac{hc}{H_f - H_i}$. If the continuous heating rate is greater than the equivalent cooling rate then
\begin{equation}
  A^{\rm cont}_{fi} = \frac{1}{\Delta H_i} \frac{dH_{\rm net}}{dt}
\end {equation}
for $f = i + 1$, where $\Delta H_i$ is the width of bin $i$ and $H_{\rm net}$ is the net heating rate $\frac{dH_{\rm heat}}{dt} - \frac{dH_{\rm cool}}{dt}$.

\subsection{Radiative cooling}

Dust grains of temperature $T$ emit radiation at wavelength $\lambda$ with intensity $Q_{\lambda} B(\lambda,T)$ where $B(\lambda,T)$ is the Planck function, causing them to lose energy. The transition rates are similar to those for absorption of a photon, with
\begin{equation}
  A_{fi} = 4 \pi^2 a^2 Q_{\lambda_{fi}} B(\lambda,T_i) \frac{hc \Delta H_f}{\left( H_i - H_f \right)^3}
\end{equation}
for $f < i$ where $T_i$ is the temperature of a grain in enthalpy bin $i$, and
\begin{equation}
  A'_{1i} = 4 \pi^2 a^2 \int_0^{\lambda_{\rm min}} \lambda Q_{\lambda} B(\lambda,T_i) d \lambda / hc
\end{equation}
where $\lambda_{\rm max} = \frac{hc}{H_i - H_{\rm min}}$. The continuous cooling rate is given by
\begin{equation}
  \frac{dH_{\rm cool}}{dt} = 4 \pi^2 a^2 \int^{\infty}_{\lambda_0} Q_{\lambda} B(\lambda,T_i) d \lambda
\end{equation}
where $\lambda_0 = \frac{hc}{H_i - H_f}$, and for $\frac{dH_{\rm cool}}{dt} > \frac{dH_{\rm heat}}{dt}$
\begin{equation}
  A^{\rm cont}_{fi} = - \frac{1}{\Delta H_i} \frac{dH_{\rm net}}{dt}
\end {equation}
for $f = i - 1$.

\subsection{Collisional heating}

As with photons, a collision between a dust grain and a particle (either an electron or an atom/ion) can result in a transfer of energy to the dust grain. Unlike with photons, a colliding particle does not necessarily transfer all its energy to the grain, and the amount of heating depends on the particle energy as well as the dust properties. The transition rate between enthalpy bins due to particle heating is given by
\begin{equation}
  A_{fi} = \pi a^2 n \int f(E) v(E) \delta(\Delta E) dE
\end{equation}
where $n$ is the number density of particles, $f(E)$ is the probability distribution of particle energies, $v(E)$ is the velocity of a particle with energy $E$, $\Delta E$ is the energy transferred to the dust grain and $\delta (\Delta E)$ is a function such that
\begin{equation}
  \delta(\Delta E) =
  \begin{cases}
    0 \quad |\Delta E - (H_f - H_i)| < \Delta H_f / 2 \\
    1 \quad {\rm otherwise}
  \end{cases}
\end{equation}
where $\Delta H_f$ is the width of enthalpy bin $f$. The additional heating rate to enthalpies higher than $H_N$ is given by
\begin{equation}
  A'_{Ni} = \pi a^2 n \int f(E) v(E) \delta ' (\Delta E) dE
\end{equation}
where
\begin{equation}
  \delta ' (\Delta E) =
  \begin{cases}
    0 \quad \Delta E < H_{\rm max} - H_i \\
    1 \quad {\rm otherwise}
  \end{cases}
\end{equation}
and the continuous heating rate is given by
\begin{equation}
  \frac{dH_{\rm heat}}{dt} = \pi a^2 n \int f(E) v(E) \delta '' (\Delta E) \Delta E dE
\end{equation}
where
\begin{equation}
  \delta '' (\Delta E) =
  \begin{cases}
    1 \quad \Delta E < H_f - H_i \\
    0 \quad {\rm otherwise}
  \end{cases}
\end{equation}
for $f = i + 1$. For electrons, the transferred energy $\Delta E$ is determined as a function of $E$ using the method described by \citet{dwek1996b}. For a dust grain of stopping thickness $R_0 = 4 a \rho / 3$ where $\rho$ is the density, if the electron range $R_1(E) \le R_0$ then $\Delta E = E$. For $R_1(E) > R_0$, $\Delta E = E - E'$ where $R(E') = R_1 - R_0$. A function for $R(E)$ based on fits to experimental data is given in \citet{dwek1996b}. For atoms and ions, \citet{dwek1987} gives the transferred energy as
\begin{equation}
  \Delta E =
  \begin{cases}
    E \quad E \le E' \\
    E' \quad E > E'
  \end{cases}
\end{equation}
where $E'$ is listed for various nuclei as a function of grain radius $a$ in \citet{dwek1987}.

\section{The gas mass for the clumped component}
\label{sec:clump}

The ejecta in Cas A is primarily composed of oxygen \citep{chevalier1979,willingale2003}. For neutral and doubly ionized oxygen, integrated line fluxes were estimated by scaling up the mean [O I] $63 \um$ and [O III] $52 \um$ fluxes (per LWS aperture) from \citet{docenko2010} by a factor of $10.5$ to account for the ratio of remnant to aperture area, giving $1.3 \times 10^{-10} \flux$ and $5.4 \times 10^{-10} \flux$ respectively. For triply ionized oxygen, the integrated [O IV] $25.9 \um$ flux from \citet{smith2009} is $4.95 \times 10^{-10} \flux$. Singly ionized oxygen, which has no IR transitions, has an [O II] $7325$\r{A} line flux of $5.93 \times 10^{-13} \flux$ in the integrated Cas A spectrum of \citet{milisavljevic2013}. The flux in the nearby [Ar III] $7136$\r{A} line is $8.13 \times 10^{-14} \flux$, while the [Ar III] $8.99 \um$ integrated flux from \citet{smith2009} is $1.24 \times 10^{-10} \flux$. The predicted ratio of $F(8.99 \um)/F(7136$\r{A}) from EQUIB \citep{howarth1981}, using an electron density of $480 \pcc$ (the mean value from the [O III] $52 \um/88 \um$ line ratios from \citet{docenko2010}) and a temperature of $10^4 \kel$, is $0.898$ - the observed $7136$\r{A} flux is therefore $1698$ times weaker than expected, corresponding to $8.07 \, {\rm mag}$ of extinction. Using the Galactic extinction law from \citet{howarth1983} with $R = 3.1$, the expected extinction at $7325$\r{A} is $7.77 \, {\rm mag}$, and the dereddened integrated [O II] $7325$\r{A} flux is therefore $7.64 \times 10^{-10} \flux$. With the same values for $\nel$ and $T$, and a distance to the SNR of $3.4 \kpc$ \citep{reed1995}, we use EQUIB to determine the number of oxygen nuclei at each stage of ionization and the corresponding masses, given in Table \ref{tab:oxy}. The total oxygen gas mass is found to be $0.49 \msun$. \citet{docenko2010} find that $N({\rm other})/N({\rm O}) = 0.10$, where $N({\rm other})$ is the number of all non-oxygen nuclei, with sulphur comprising half of $N({\rm other})$. Assuming a mass per `other' nuclei of $32 \mh$, the ratio $M({\rm other})/M({\rm O})$ is $0.20$, and the total gas mass of the optical/IR-emitting clumps is $0.59 \msun$.

\begin{table*}
  \centering
  \caption{Deduced number of oxygen nuclei in various ionization states, and corresponding masses, in the optical/IR-emitting gas component in the Cas A SNR.}
  \begin{tabular}{cccccc}
    \hline
    Ion stage & O$^0$ & O$^+$ & O$^{2+}$ & O$^{3+}$ & Total \\
    \hline
    Number & $5.29 \times 10^{54}$ & $2.91 \times 10^{55}$ & $1.74 \times 10^{54}$ & $2.01 \times 10^{53}$ & $3.63 \times 10^{55}$ \\
    Mass / $\msun$ & $0.07$ & $0.39$ & $0.02$ & $0.003$ & $0.49$ \\
    \hline
  \end{tabular}
  \label{tab:oxy}
\end{table*}

%%%%%%%%%%%%%%%%%%%%%%%%%%%%%%%%%%%%%%%%%%%%%%%%%%

% Don't change these lines
\bsp	% typesetting comment
\label{lastpage}
\end{document}